\documentclass[preprint,12pt]{aastex}
\newcommand{\fracb}[2]{\left(\frac{#1}{#2}\right)}
\newcommand{\mean}[1]{\langle{#1}\rangle}

\begin{document}

\title{The effects of sub-shells in highly magnetized relativistic flows}

\author{Jonathan Granot\altaffilmark{1,2,3}}

\altaffiltext{1}{Racah Institute of Physics, The Hebrew University, Jerusalem 91904, Israel}
\altaffiltext{2}{Raymond and Beverly Sackler School of Physics \& Astronomy, Tel Aviv University, Tel Aviv 69978, Israel}
\altaffiltext{3}{Centre for Astrophysics Research, University of Hertfordshire, College Lane, Hatfield, AL10 9AB, UK; j.granot@herts.ac.uk}

\begin{abstract}

Astrophysical sources of relativistic jets or outflows, such as
gamma-ray bursts (GRBs), active galactic nuclei (AGN) or
micro-quasars, often show strong time variability. Despite such
impulsive behavior, most models of these sources assume a steady state
for simplicity. Here I consider a time-dependent outflow that is
initially highly magnetized and divided into many well-separated
sub-shells, as it experiences impulsive magnetic acceleration and
interacts with the external medium. In AGN the deceleration by the
external medium is usually unimportant and most of the initial
magnetic energy is naturally converted into kinetic energy, leading to
efficient dissipation in internal shocks as the sub-shells
collide. Such efficient low-magnetization internal shocks can also
naturally occur in GRBs, where the deceleration by the external medium
can be important. A strong low-magnetization reverse shock can
develop, and the initial division into sub-shells allows it to be
relativistic and its emission to peak on the timescale of the prompt
GRB duration (which is not possible for a single shell). Sub-shells
also enable the outflow to reach much higher Lorentz factors that help
satisfy existing constraints on GRBs from intrinsic pair opacity and
from the afterglow onset time.
 
\end{abstract}


\section{Introduction}
\label{sec:introduction}

The composition and acceleration mechanism of the relativistic
outflows that power gamma-ray bursts (GRBs) are important open
questions in this field~\citep[for a review see][]{Piran05}. In
particular, their degree of magnetization and the role of magnetic
fields in the acceleration or collimation of GRB outflows is of great
interest. In recent years, models in which the outflow is highly
magnetized close to the source, and possibly also at very large
distances from the source where much of the observed emission is
produced, have been gaining
popularity~\citep{DS02,VK03,LB03,GS06,Kom07,Tchek10,Lyub10b} and may
provide a viable alternative to the traditional fireball model.

Some other relativistic outflow sources are even more likely strongly
magnetized near the central source. Pulsar winds are almost certainly
Poynting flux dominated near the source, and the same very likely also
holds for active galactic nuclei (AGN) and tidal disruption events
(TDEs) of a star by a super-massive black hole. In AGN and TDEs, which
are often highly variable (i.e. impulsive) suggesting sub-shells in
the outflow, since the central accreting black hole is super-massive
then even close to it the Thompson optical depth $\tau_T$ may not be
high enough for thermal acceleration by radiation pressure -- the main
competition to magnetic acceleration -- to work efficiently
\citep[e.g.,][]{Ghis11}. Observations of relevant sources, such AGN,
GRBs or pulsar wind nebulae suggest that the outflow magnetization is
rather low at large distances from the source. An important
outstanding question concerning outflows that start off highly
magnetized near the source is how they convert most of their initial
electromagnetic energy into the energy in the bulk and random motions
particles, where the latter also produces the radiation we observe
from these sources. This is known as the $\sigma$ problem, namely how
to transform from $\sigma\gg 1$ near the source to $\sigma\ll 1$ very
far from the source, where the magnetization parameter $\sigma$ is the
Poynting-to-matter energy flux ratio.  Different approaches to this
problem have been considered so far.

Outflows that are initially Poynting flux dominated are usually
treated (for simplicity) under ideal MHD, axi-symmetry and
steady-state. Under these conditions, however, it is difficult to
achieve sufficiently low magnetization ($\sigma < 1$ or even
$\sigma\ll 1$) at large distances from the source that would allow
efficient dissipation in internal
shocks~\citep{Kom09,Lyub09,Lyub10a}. A possible solution is that the
magnetization remains high ($\sigma\gg 1$) also at large distances
from the source and the observed emission is powered by magnetic
reconnection rather than by internal
shocks~\citep{LB03,Lyut06}. Alternatively, the non-axi-symmetric kink
instability could randomize the direction of the magnetic field,
making it behave more like a fluid and enhancing magnetic
reconnection, which both increase the acceleration and help lower the
magnetization~\citep{HB00,DS02,GS06}. Another option that may be
relevant for AGN and GRBs~\citep{Lyub10b}, is that if the Poynting
flux dominated outflow has alternating fields (e.g. a striped wind)
then the Kruskal-Schwarzschild instability (i.e. the magnetic version
of the Rayleigh-Taylor instability) of the current sheets could lead
to significant magnetic reconnection, which in turn increases the
initial acceleration resulting in a positive feedback and
self-sustained acceleration that leads to a low magnetization.

Here I focus on the effects of strong time dependence -- impulsive
outflows that are initially highly magnetized, with $\sigma_0\gg 1$,
under ideal MHD. \citet[][hereafter paper I]{GKS11} have recently
found a new impulsive magnetic acceleration mechanism for relativistic
outflows, which is qualitatively different from its Newtonian
analog~\citep{Cont95}, and can lead to kinetic energy dominance and
low $\sigma$ that allow for efficient dissipation in internal
shocks. Paper I focused mainly on the acceleration of an initially
highly magnetized shell of plasma into vacuum. The initial magnetic
energy can be almost fully converted into kinetic form as the shell
expands radially under its own magnetic pressure. Initially, while it
is still highly magnetized, this expansion leads to impulsive
acceleration and the resulting increase in the Lorentz contraction
almost exactly cancels the increase in the shell's width in its own
rest frame, leading to a constant width in the lab (i.e. central
source) frame. Once it becomes kinetically dominated it starts
spreading radially significantly also in the lab frame, and its
magnetization quickly drops to $\sigma\ll 1$.

Paper I only briefly discussed the effects of the interaction with the
external medium.  The interaction with an unmagnetized external medium
whose density varies as a power-law with the distance from the central
source is analyzed in detail in an accompanying
paper~\citep[][hereafter paper II]{Granot11}.  The present work
generalizes this self-consistent treatment of the combined impulsive
magnetic acceleration and deceleration by the external medium, by
examining the effects of an outflow that is initially divided into
many well separated sub-shells, instead of a single shell. Such
multiple sub-shells are both naturally expected in highly variable
sources, and are also required in order to produce internal shocks
(where enabling efficient internal shocks is one of the main
motivations for this type of model).

The basic test case examined in paper I was a highly magnetized shell
initially at rest, whose back leans against a conducting wall and
whose front edge faces vacuum, while in paper II the vacuum is
replaced by an unmagnetized external medium. As discussed in paper I,
while such a setup might be directly applicable for, e.g, a giant
flare from a soft gamma repeater, for most relevant astrophysical
sources (such as GRBs, AGN or micro-quasars) we expect that initially
a quasi-steady acceleration takes place and saturates at some radius
(i.e. distance from the source), while the impulsive acceleration
mechanism takes over at a larger radius. The dynamics most relevant
for our purposes occur after the impulsive acceleration takes over,
and are insensitive to whether part of the earlier acceleration
occurred in the quasi-steady regime (while the jet was being
collimated). However, during the impulsive acceleration phase the
outflow still retains memory of its initial properties when it was
ejected from the central source, namely the time history of its
initial magnetization $\sigma_0$ and ejected energy per unit time,
which is reflected in the isotropic equivalent luminosity $L$
(i.e. energy flux through a sphere of fixed radius if the outflow
occupied all of the solid angle). During this phase the
outflow opening angle remains roughly constant, and its dynamics
are essentially spherical, and equivalent to those for the planar
case, as shown in paper I.

The sub-shells are assumed to be initially well separated, uniform
with sharp edges and separations comparable to or larger than their
widths, and with a large contrast (e.g. in energy density) between the
sub-shells and the ``gaps'' between them (as might be the result of
large variations in the energy output rate from the central source
into the outflow). A modest contrast and/or very smooth edges for the
sub-shells might cause the sub-shells to collide and merge earlier on,
while they are still highly magnetized, making the outflow
subsequently behave closer to the steady regime than to the impulsive
regime. Quantifying these effects, however, is beyond the scope of
this paper and is left for a separate work.

The basic physical setup explored in this work is described
in~\S~\ref{sec:sub-shells}, while~\S~\ref{sec:identical} studies in
detail the main test case, namely the dynamics of $N\gg 1$ identical
sub-shells. I find several different cases for the dynamics, which
naturally divide into three groups: (i) a low magnetization ``thin
shell'' (cases 1 and 2$^*$) where a strong mildly relativistic reverse
shock develops on a timescale larger than the duration of the prompt
GRB emission ($T_{\rm dec}\gg T_{\rm GRB}$), (ii) a low or mild
magnetization ``thick shell'' (cases 2 and 3A), where a relativistic
reverse shock develops whose emission peaks on a timescale comparable
to that of the prompt GRB emission ($T_{\rm dec}\sim T_{\rm GRB}$),
(iii) a high magnetization ``thick shell'' (case 3B), where the
reverse shock and its associated emission are strongly suppressed
(with $T_{\rm dec}\sim T_{\rm GRB}$ where $T_{\rm dec}$ in this case
can be identified through the afterglow emission, which peaks on this
timescale).  The first two groups, (i) and (ii), naturally produce
internal shocks at mild magnetization and high Lorentz factors, which
can help accommodate GRB observations.
The effects of varying the initial magnetization between different
sub-shells are explored in \S~\ref{sec:vary-sigma}, and I show
that this is not expected to strongly affect the main results, even
for strong initial variations (and the same holds for order unity
variations in the other model parameters).  The new results found in
this work are summarized in \S~\ref{sec:sum} and their implications
are discussed in \S~\ref{sec:dis}.

\section{Sub-shells versus a single shell}
\label{sec:sub-shells}

Here the results of paper II for a single spherical shell of initial
(lab-frame) width $\Delta_0 \approx R_0$, energy $E$, luminosity
$L\approx Ec/\Delta_0$ and initial magnetization $\sigma_0 =
B_0^2/(4\pi\rho_0c^2)\gg 1$ (with a magnetic field normal to the
radial direction, where $B_0$ and $\rho_0$ are its initial magnetic
field and rest-mass density, respectively), are generalized to the
case where the same initially uniform shell is divided into $N\gg 1$
identical sub-shells of initial widths $\Delta_{\rm 0,sh} =\Delta_0/N$
and separations $\Delta_{\rm gap}\gtrsim \Delta_{\rm 0,sh}$, energy
$E_{\rm sh} = E/N$, with the same luminosity ($L_{\rm sh} = E_{\rm
sh}c/\Delta_{\rm 0,sh} = Ec/\Delta_0= L$) and initial magnetization
($\sigma_{\rm 0,sh} =\sigma_0$) as the original single shell.  The
total initial width of all the sub-shells including the gaps between
them is $\Delta_{\rm tot} \approx (\Delta_{\rm 0,sh}+\Delta_{\rm
gap})N =\Delta_0\tilde{\Delta}$ where $\tilde{\Delta}\equiv
1+\Delta_{\rm gap}/\Delta_{\rm 0,sh}$.

The unperturbed external medium is taken to be cold, unmagnetized, and
with a rest mass density that varies as a power-law (of index $k$)
with the distance $R$ from the central source, $\rho_1 = AR^{-k}$. The
external medium interacts directly only with the first (i.e. leading)
sub-shell, which sweeps it up and drives a strong relativistic shock
into it, where a contact discontinuity (CD) separates the shocked
external medium and the material in the first sub-shell. The
subsequent (or trailing) sub-shells do not directly interact with the
external medium. Instead, they interact directly only with their
neighboring sub-shells (i.e., each sub-shell interacts only with the
sub-shell/s just in front of it and/or just behind it),
and initially they propagate in the relatively evacuated region left
behind by the preceding sub-shell.

Since the sub-shells are assumed to be initially well separated, each
sub-shell starts accelerating independently. The first (or leading)
sub-shell interacts with the external medium, and until the second
sub-shell collides with it from behind its dynamics essentially follow
those of a single (albeit, relatively narrow) shell, which were studied
in detail in paper II. Following the results of paper II (and adding
`sh' to the subscripts of quantities that refer to a single sub-shell
rather than to the whole outflow), it has the following critical
radii:
\begin{equation}
R_{\rm 0,sh}\sim \frac{R_0}{N}\ ,\quad\quad
R_{\rm c,sh}\sim \frac{R_c}{N}\ ,\quad\quad
R_{u,{\rm sh}}\sim R_u N^{-4/(10-3k)}\ ,\quad\quad
R_{\rm cr,sh}\sim R_{\rm cr}N^{-2/(4-k)}\ .
\end{equation}
Here $R_{\rm 0,sh}$ ($R_0$) is the initial radius of the sub-shell
(or the original single shell), essentially equal to its initial width,
$\Delta_{\rm 0,sh}$ ($\Delta_0$), where its typical Lorentz factor and
magnetization are $\mean{\Gamma_{\rm sh}}\sim\sigma_{\rm 0,sh}^{1/3}$
($\mean{\Gamma}\sim\sigma_0^{1/3}$) and $\mean{\sigma_{\rm sh}}\sim
\sigma_{\rm 0,sh}^{2/3}$ ($\mean{\sigma}\sim\sigma_0^{2/3}$),
respectively, and in regimes I or II from paper II at $R>R_{\rm 0,sh}$
($R>R_0$) they start to evolve as $\mean{\Gamma_{\rm
sh}}\sim\sigma_0/\mean{\sigma_{\rm sh}} \sim (\sigma_{\rm
0,sh}R/R_{\rm 0,sh})^{1/3}$
($\mean{\Gamma}\sim\sigma_0/\mean{\sigma}\sim(\sigma_0R/R_0)^{1/3}$);
$R_{\rm c,sh}$ ($R_c$) is the coasting radius where if the external
density is low enough, corresponding to regime I from paper II, the
sub-shell (or the original single shell) becomes kinetically dominated
and starts to coast at $\mean{\Gamma_{\rm sh}}\sim\sigma_{\rm 0,sh}$
($\mean{\Gamma}\sim\sigma_0$) while its magnetization rapidly drops
with radius, $\mean{\sigma_{\rm sh}}(R>R_{\rm c,sh})\sim R_{\rm
c,sh}/R$ ($\mean{\sigma}(R>R_c)\sim R_c/R$); $R_{u,{\rm sh}}$ ($R_u$)
is the radius where, in regime II from paper II, the typical Lorentz
factor of the sub-shell (or the original single shell) becomes similar
to that just behind the CD, stops growing as $R^{1/3}$, and instead
starts evolving as $\mean{\Gamma_{\rm sh}}\sim
\sigma_0/\mean{\sigma_{\rm sh}}\sim \Gamma_{\rm cr,sh}(R/R_{\rm
cr,sh})^{(k-2)/4}$ ($\mean{\Gamma}\sim
\sigma_0/\mean{\sigma}\sim\Gamma_{\rm cr}(R/R_{\rm cr})^{(k-2)/4}$)
up to the radius $R_{\rm cr,sh}$ ($R_{\rm cr}$);
$R_{\rm cr,sh} \sim R_{\rm 0,sh}\Gamma_{\rm cr,sh}^2$ ($R_{\rm cr}
\sim R_0\Gamma_{\rm cr}^2$) is the deceleration radius in regime II
(from paper II) where most of the energy originally in the sub-shell
(or the original single shell) is transfered to the shocked external
medium.

The critical Lorentz factors, $\Gamma_{\rm cr,sh}$ or $\Gamma_{\rm
cr}$, which signify the borderline between regimes I ($\sigma_{\rm
0,sh} < \Gamma_{\rm cr,sh}$ or $\sigma_0<\Gamma_{\rm cr}$) and II
($\Gamma_{\rm cr,sh} <\sigma_{\rm 0,sh} < \Gamma_{\rm
cr,sh}^{(12-3k)/2}$ or $\Gamma_{\rm cr} < \sigma_0 < \Gamma_{\rm
cr}^{(12-3k)/2}$), satisfy
\begin{equation}
\frac{\Gamma_{\rm cr,sh}}{\Gamma_{\rm cr}} \sim N^{(2-k)/(8-2k)}
\sim \fracb{R_{\rm cr,sh}}{R_{\rm cr}}^{(k-2)/4}\ .
\end{equation}
This essentially reflects the fact that since the sub-shell and the
corresponding (or original) single shell have the same luminosity,
they share the same line in the $\Gamma\,$--$\,R$ plane corresponding to a
pressure balance at the CD between the bulk of the shell (or
sub-shell) and the shocked external medium (for details see paper II),
$\Gamma_{\rm CD} \sim (L/Ac^3)^{1/4}R^{(k-2)/4}
\sim\Gamma_{\rm cr}(R/R_{\rm cr})^{(k-2)/4}\sim {}\Gamma_{\rm
cr,sh}(R/R_{\rm cr,sh})^{(k-2)/4}$.
When comparing to observational constraints, the numerical value of
$\Gamma_{\rm cr}$ is very useful,
\begin{eqnarray}\nonumber
\Gamma_{\rm cr} &=& \left[\frac{(3-k)E}{4\pi Ac^2\Delta_0^{3-k}}\right]^\frac{1}{8-2k} = 
\left[\frac{(3-k)(1+z)^{3-k}\tilde{\Delta}^{3-k}E}{4\pi Ac^{5-k}T_{\rm GRB}^{3-k}}\right]^\frac{1}{8-2k}
\\ \nonumber
\\ \label{eq:Gamma_cr}
&=&\left\{\matrix{
395\,\zeta^{3/8}\tilde{\Delta}^{3/8}E_{53}^{1/8}n_0^{-1/8}T_{30}^{-3/8}
& (k=0)\ , \cr & \cr
88\,\zeta^{1/4}\tilde{\Delta}^{1/4}E_{53}^{1/4}A_*^{-1/4}T_{30}^{-1/4}
& (k=2)\ , }\right.
\end{eqnarray}
(paper II) where $\zeta=(1+z)/3$, $z$ is the source redshift, $T_{\rm
GRB}= (1+z)\Delta_{\rm tot}/c=30T_{30}\;$s is the observed duration of
the GRB and $E=10^{53}E_{53}\;$erg is the total (isotropic equivalent)
energy of the ejecta. Numerical values are provided for the physically
interesting cases of $k=0$ (a uniform medium of number density $n =
A/m_p = n_0\;{\rm cm^{-3}}$), and $k = 2$ (corresponding to the
stellar wind of a massive star progenitor, with $A = 5\times
10^{11}A_*\;{\rm gr\; cm^{-1}}$). The subsequent sections largely
follow paper II, build upon it and use consistent notations.

\section{Identical sub-shells: dynamical regimes for different values of 
$k$ and $\sigma_{\rm 0,sh}$}
\label{sec:identical}

\subsection{Moderate (or no) external density stratification: $k<2$}
\label{sec:k<2}

\begin{figure}
\centerline{\includegraphics[height=5.9cm]{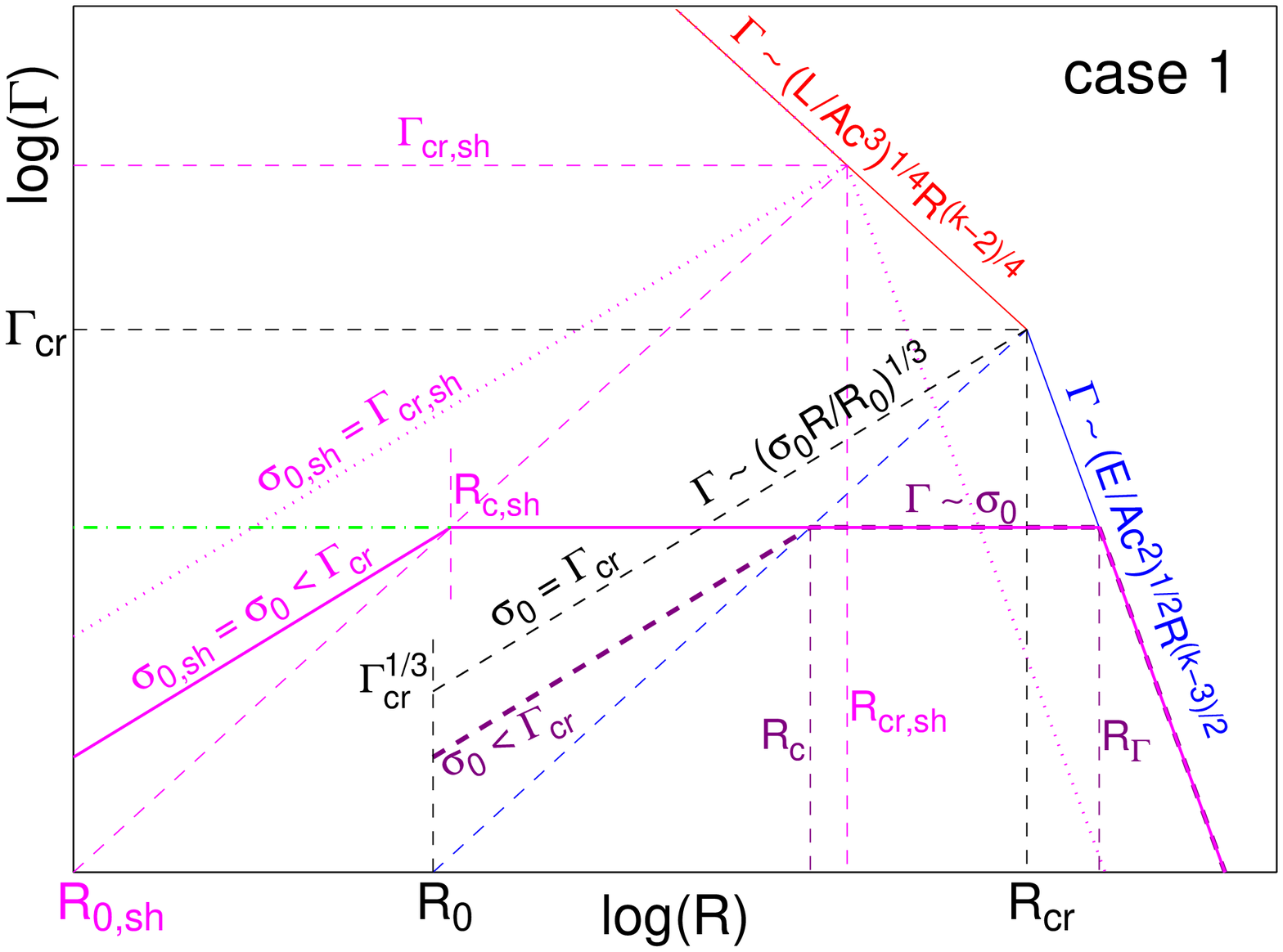}}
\vspace{0.5cm}
\centerline{\includegraphics[height=5.9cm]{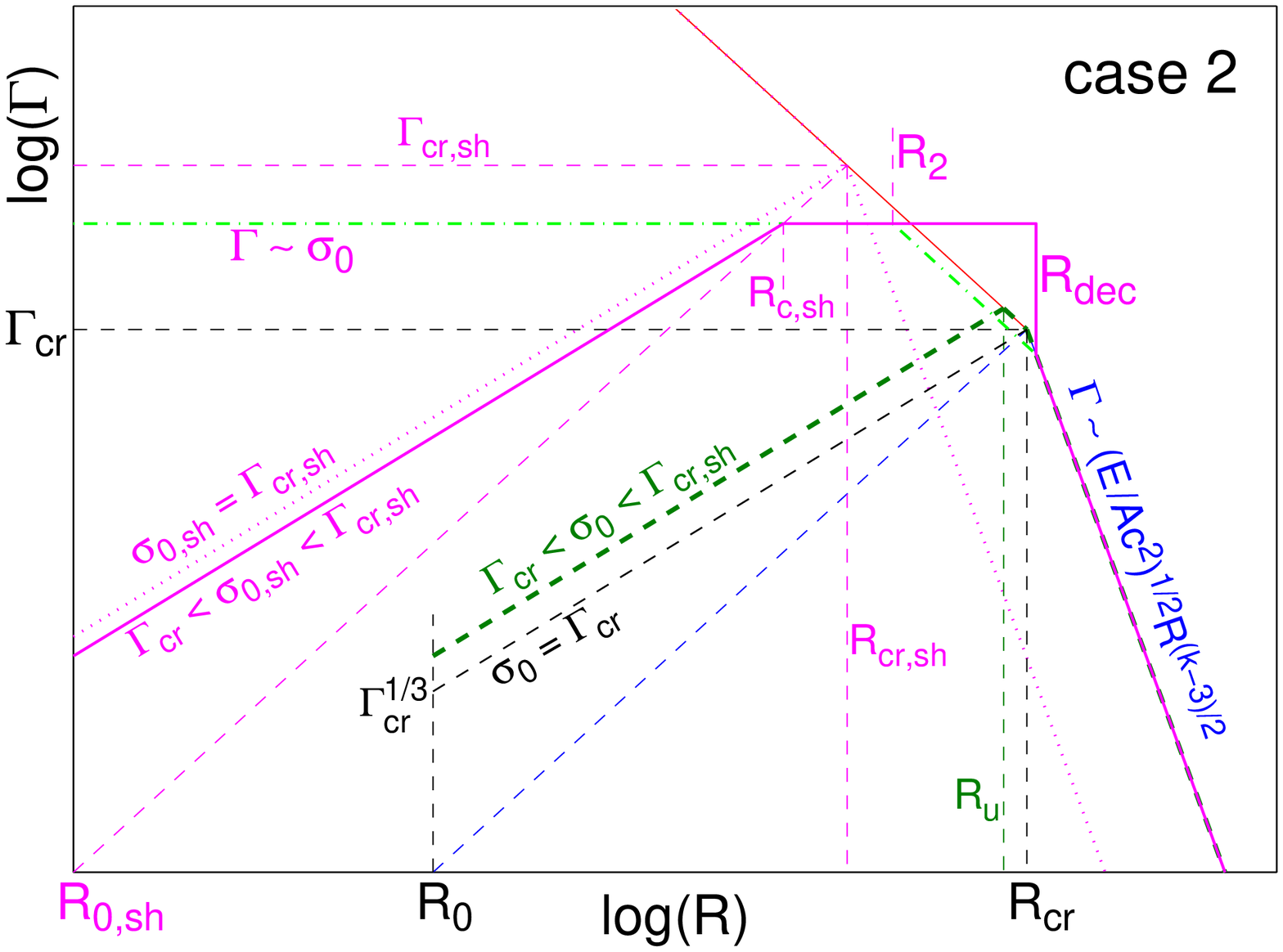}}
\vspace{0.5cm}
\centerline{\includegraphics[height=5.9cm]{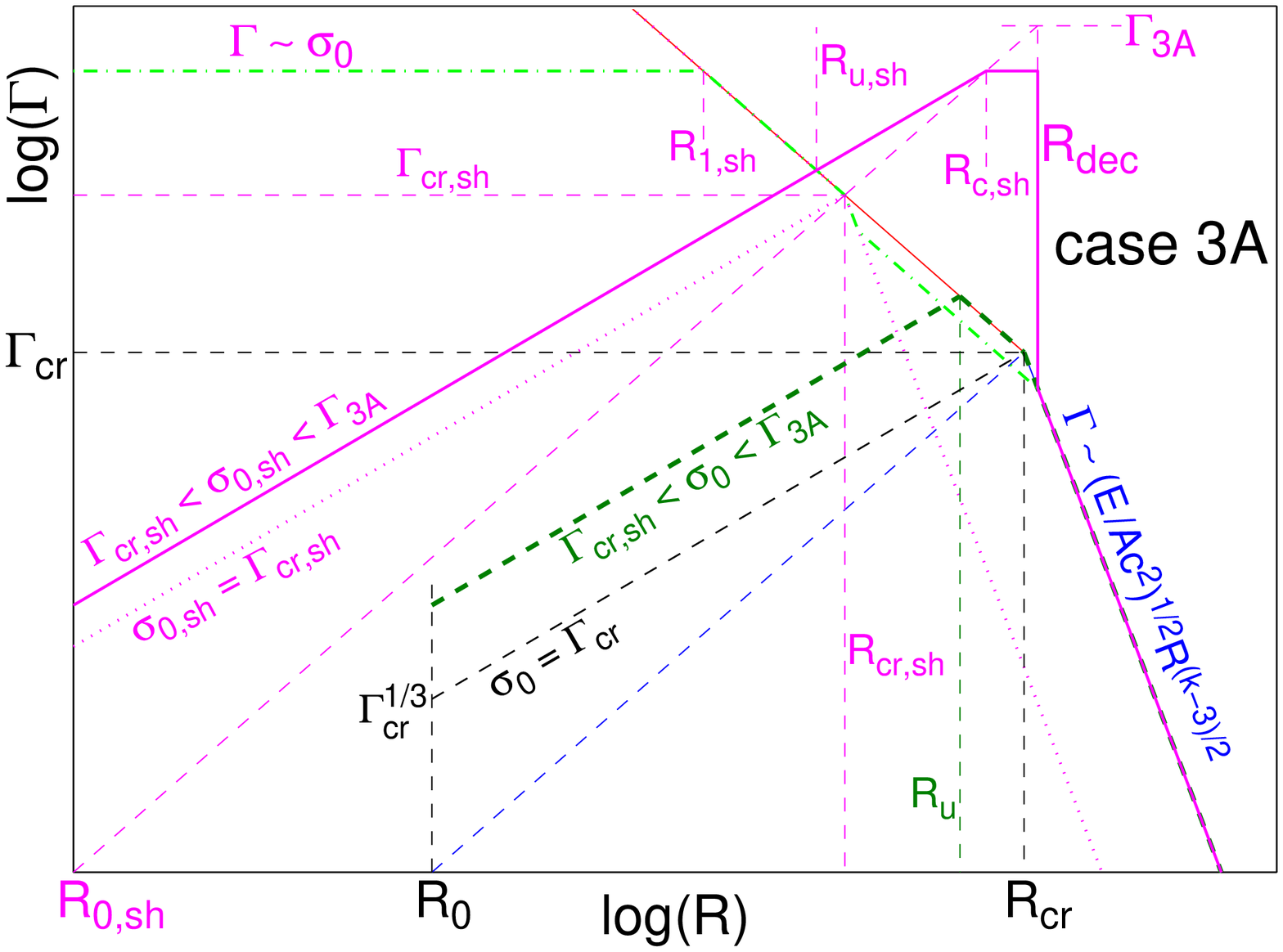}}
\caption{
The evolution of the typical or energy-weighted mean Lorentz factor
($\mean{\Gamma}$; {\it thick solid magenta line}) and the Lorentz
factor of the CD ($\Gamma_{\rm CD}$; {\it thick dashed-dotted green
line}) with radius, $R$, for $N\gg 1$ equal, initially highly
magnetized ($\sigma_{\rm 0,sh}\gg 1$) sub-shells, compared to a single
uniform shell ({\it thick dashed line}; purple in the top panel
corresponding to regime I, and dark green in the middle or bottom
panels corresponding to regime II), with the same initial
magnetization ($\sigma_0=\sigma_{\rm 0,sh}$) and luminosity (or energy
density), as well as the same total energy and net width (not counting
the initial gaps between the sub-shells), for $k<2$.  }
\label{fig:sub-shells_k<2}
\end{figure}

\noindent{\bf Case 1:}
If $\sigma_{\rm 0,sh} = \sigma_0 < \Gamma_{\rm cr}$ (or $R_{\rm c,sh}<
R_{\rm cr}/N$; top panel of Fig.~\ref{fig:sub-shells_k<2}) then even
the first (leading) sub-shell reaches its coasting radius $R_{\rm
c,sh}$ without being significantly affected by the external
medium. Since the subsequent (trailing) sub-shells start propagating
in the evacuated region behind the first shell, they are initially not
affected by the external medium. Thus, all of the sub-shells
accelerate largely independently until becoming kinetically dominated
at $R_{\rm c,sh}\sim R_{\rm 0,sh}\sigma_{\rm 0,sh}^2 \sim R_c/N$,
where they start coasting (at $\mean{\Gamma_{\rm sh}}\sim\sigma_{\rm 0,sh}$) and
spreading radially. Then, they soon collide with each other and form
internal shocks at $R_{\rm IS}\sim (\Delta_{\rm gap}/\Delta_{\rm
0,sh})R_{\rm c,sh}$. After the internal shocks subside, the resulting
merged shell of width $\sim\Delta_{\rm
tot}\approx\Delta_0\tilde{\Delta}\sim\Delta_0(\Delta_{\rm
gap}/\Delta_{\rm 0,sh})$ starts spreading radially only at $R_s\sim
\Delta_{\rm tot}\sigma_{\rm 0,sh}^2\sim (\Delta_{\rm gap}/\Delta_{\rm
0,sh})R_c\sim NR_{\rm IS}$ where its mean magnetization is still
similar to that near the internal shocks radius, $\mean{\sigma}(R_s)
\sim \mean{\sigma}(R_{\rm IS})\sim\Delta_{\rm 0,sh}/\Delta_{\rm
gap}\lesssim 1$. At $R>R_s$ it starts evolving as $\Delta \sim
(R/R_s)\Delta_{\rm tot} \sim (R/R_c)\Delta_0$ and $\mean{\sigma} \sim
(R_s/R)\mean{\sigma}(R_s)\sim R_c/R$, similar to the single wide shell
that corresponds to the sub-shells considered here if they had no
initial gaps between them.  Thus, the merged shell similarly follows
the unmagnetized ``thin shell'' case (regime I from paper II), and the
reverse shock that forms becomes mildly relativistic near the
deceleration radius, $R_{\rm dec}\sim R_\Gamma\sim
(E/\sigma_0^2Ac^2)^{1/(3-k)}\sim R_c(\Gamma_{\rm
cr}/\sigma_0)^{(8-2k)/(3-k)}$ (here $R_\Gamma$ is the radius at which
a rest mass $E/\Gamma_0^2c^2$ of the external medium is swept up; see
Eq.~[4] of paper II), where it finishes crossing the merged shell and
most of the energy is transfered to the shocked external medium. The
magnetization at this radius is already low, $\mean{\sigma}(R_{\rm
dec}) \sim (\sigma_0/\Gamma_{\rm cr})^{(8-2k)/(3-k)}\sim R_c/R_{\rm
dec}\ll 1$. At $R>R_{\rm dec}$ the flow quickly approaches the
\citet{BM76} self-similar solution, where $\mean{\Gamma} \sim
(E/Ac^2)^{1/2}R^{(k-3)/2} \sim \sigma_0(R/R_\Gamma)^{(k-3)/2}$. The
maximal Lorentz factor reached by the outflow in case 1 is
$\Gamma\sim\sigma_0 < \Gamma_{\rm cr}$.

\noindent{\bf Case 2:}
If $\Gamma_{\rm cr} < \sigma_0=\sigma_{\rm 0,sh} < \Gamma_{\rm cr,sh}$
(i.e. $1<\sigma_0/\Gamma_{\rm cr}<N^{(2-k)/(8-2k)}$) or $R_{\rm cr}/N
< R_{c,{\rm sh}} < R_{\rm cr,sh}\sim R_{\rm cr}N^{-2/(4-k)}$ (middle
panel of Fig.~\ref{fig:sub-shells_k<2}), then similarly to case 1
above all of the sub-shells accelerate independently until becoming
kinetically dominated at $R_{\rm c,sh}$, where they start to coast (at
$\mean{\Gamma_{\rm sh}}\sim\sigma_{\rm 0,sh}$) and spread radially,
and soon collide with each other at $R_{\rm IS}\sim (\Delta_{\rm
gap}/\Delta_{\rm 0,sh})R_{\rm c,sh}$. The main difference is that in
this case the resulting merged shell is decelerated significantly by
the external medium before it starts to spread radially\footnote{This
follows from the relations $R_s/R_2 > R_s/R_{\rm dec} \sim
\tilde{\Delta}^{(3-k)/(4-k)}R_c/R_{\rm cr}
{}\sim\tilde{\Delta}^{(3-k)/(4-k)}(\sigma_0/\Gamma_{\rm cr})^2 > 1$.},
while its typical magnetization is modest, $\sim \mean{\sigma}(R_{\rm
IS})\sim\Delta_{\rm 0,sh}/\Delta_{\rm gap}\lesssim 1$, and therefore a
strong, highly relativistic reverse shock develops. Hence, case 2
effectively reverts to the unmagnetized (or at most mildly magnetized)
``thick shell'' case, where a bright reverse shock emission on a
timescale comparable to that of the prompt GRB emission ($T_{\rm
dec}\sim T_{\rm GRB}$) is expected. Note that both the observed
deceleration time, $T_{\rm dec} \sim R_{\rm
dec}/c\mean{\Gamma}^2(R_{\rm dec})$ (i.e. the timescale on which both
the reverse shock and afterglow emission peak) and $T_{\rm GRB}$ scale
linearly with $\tilde{\Delta}$ in this case so that their ratio (or
the relation $T_{\rm dec}\sim T_{\rm GRB}$) is independent of
$\tilde{\Delta}$. The effective luminosity of the merged shell is
somewhat lower than that of the original sub-shells or the
corresponding single wide shell, $L_{\rm merged}\approx Ec/\Delta_{\rm
tot} = L/\tilde{\Delta}$, and therefore a strong relativistic reverse
shock develops at the radius $R_2$ where the Lorentz factor of the CD
at larger radii, $\Gamma_{\rm CD}(R_2 < R < R_{\rm dec}) \sim (L_{\rm
merged}/Ac^3)^{1/4}R^{(k-2)/4}$, becomes comparable to the coasting
Lorentz factor, $\Gamma_{\rm CD}(R<R_2)\sim
\mean{\Gamma}(R_{\rm c,sh}<R<R_{\rm dec})\sim\sigma_{\rm 0,sh}$. This
implies $R_2 \sim
\tilde{\Delta}^{-1/(2-k)}R_1$ where\footnote{The reason why $R_2$ is
so close to $R_1$, up to a factor of $\sim (L_{\rm
merged}/L)^{1/(2-k)}\sim \tilde{\Delta}^{-1/(2-k)}$, is that in both
cases it is basically the radius where the Lorentz factor of the
contact discontinuity (CD), $\Gamma_{\rm CD} \sim
(L/Ac^3)^{1/4}R^{(k-2)/4}$, which is determined by pressure balance at
the CD between the shocked external medium and the bulk of the shell
(where the latter is provided by magnetic pressure for $R_1$ in regime
II and by thermal pressure behind the mildly relativistic reverse
shock for $R_2$), is $\sim \sigma_0 = \sigma_{\rm 0,sh}$.} $R_1$ is
the radius at which $\sigma = 1$ just behind the CD (for the
corresponding single wide shell) and it is given by $R_1/R_c\sim
(\Gamma_{\rm cr}/\sigma_0)^{(8-2k)/(2-k)}$ (see Eqs.~[27] and [34] in
paper II). The relativistic reverse shock finishes crossing the shell
at the deceleration radius, $R_{\rm
dec}\sim\tilde{\Delta}^{1/(4-k)}R_{\rm cr}$, where most of the energy
is transfered to the shocked external medium. This is the reason
behind the sharp drop in the energy weighted mean Lorentz factor of
the flow, $\mean{\Gamma}$, near $R_{\rm dec}$ (see middle panel of
Fig.~\ref{fig:sub-shells_k<2}) since essentially on a single dynamical
time it changes from being dominated by the coasting (unshocked) part
of the merged shell with $\Gamma\sim\sigma_{\rm 0,sh}$ to being
dominated by the shocked external medium with $\Gamma(R_{\rm
dec})\sim\Gamma_{\rm cr}\tilde{\Delta}^{-(3-k)/(8-2k)}\ll\sigma_{\rm
0,sh}$ (where near the transition, at $R\sim R_{\rm dec}$, the shocked
part of the merged shell, which has a Lorentz factor similar to that
of the shocked external medium, also holds a good fraction of the
total energy). At $R>R_{\rm dec}$ the flow quickly approaches
the~\citet{BM76} self-similar solution.  The maximal Lorentz factor
reached by the outflow in case 2 is $\mean{\Gamma}\sim\sigma_0$, and can
approach $\Gamma_{\rm cr,sh}\sim\Gamma_{\rm cr}N^{(2-k)/(8-2k)}
>\Gamma_{\rm cr}$.

\noindent{\bf Case 3:}
If $\sigma_{\rm 0,sh} = \sigma_0 > \Gamma_{\rm cr,sh}$
(i.e. $\sigma_0/\Gamma_{\rm cr}>N^{(2-k)/(8-2k)}$) or $R_{\rm c,sh} >
R_{\rm cr,sh}\sim R_{\rm cr}N^{-2/(4-k)}$ then the first sub-shell
starts to decelerate significantly (because of the $PdV$ work that it
performs across the CD on the shocked external medium) at $R_{u,{\rm
sh}}\sim R_u N^{-4/(10-3k)} \sim R_{\rm cr,sh}(\Gamma_{\rm
cr,sh}/\sigma_{\rm 0,sh})^{4/(10-3k)}$, while it is still highly
magnetized ($\mean{\sigma_{\rm sh}}(R_{u,{\rm sh}})\sim
\sigma_{\rm 0,sh}/\Gamma_{\rm CD}(R_{u,{\rm sh}})
\sim(\sigma_{\rm 0,sh}/\Gamma_{\rm cr,sh})^{(8-2k)/(10-3k)} > 1$) and
before spreading radially appreciably, thus effectively following
regime II (from paper II). The second sub-shell, however, would
effectively collide and merge with the first sub-shell only at a
radius $R_{\rm col,1}\sim (\Delta_{\rm gap}/\Delta_{\rm
0,sh})^{2/(4-k)}R_{\rm cr,sh}$ (see Appendix~\ref{app:R_col}), and
until that radius the bulk of this sub-shell\footnote{The front part
of the sub-shell would start interacting with the tail of the
preceding sub-shell at smaller radii, but the interaction would start
to significantly affect the bulk of the second sub-shell only at $R
\sim R_{\rm col, 1}$. By $R_{\rm col,1}$ the first sub-shell has
already expanded significantly in the lab frame, by a factor of
$\sim\tilde{\Delta}$, and therefore its electromagnetic energy is
reduced by the same factor (since it scales as $E_{\rm EM} \propto
B^2\Delta \propto 1/\Delta$), however its magnetization is still high
so that most of its energy is still in magnetic form and this
represents mainly the loss of energy to the shocked external medium
due to the work it performs on it at the CD.} would still accelerate
almost as if into vacuum, following $\mean{\Gamma_{\rm
sh}}\sim\sigma_{\rm 0,sh}/\mean{\sigma_{\rm sh}}\sim (\sigma_{\rm
0,sh}R/R_{\rm 0,sh})^{1/3}$. The same holds for subsequent collisions,
where the radius of the $n^{\rm th}$ collision is $R_{{\rm col,}n}
\sim (n\tilde{\Delta}/N)^{2/(4-k)}R_{\rm cr}\sim (n/N)^{2/(4-k)}R_{\rm
dec}$ (see Appendix~\ref{app:R_col}), as long as $R_{{\rm
col,}n}<R_{\rm c,sh}$ or equivalently\footnote{If $R_{{\rm col,}n} >
R_{\rm c,sh}$ or $n/N \gtrsim (\sigma_0/\Gamma_{\rm 3B})^{4-k}$, then
the sub-shell would spread radially and collide at $R_{\rm IS}\sim
\tilde{\Delta}R_{\rm c,sh}$, and would eventually be decelerated by
the relativistic reverse shock at $R \sim R_{{\rm col,}n}$ (which
might in that sense still be considered as an effective ``collision''
radius).} $n/N < (\sigma_0/\Gamma_{\rm 3B})^{4-k}$ (which is always
satisfied in case 3B below, but not in case 3A), where $\Gamma_{\rm
3B}$ is defined below. In such highly magnetized collisions the
$(n+1)^{\rm th}$ sub-shell that catches up from behind in the $n^{\rm
th}$ collision accelerates up to the time of that collision to
$\mean{\Gamma_{{\rm sh,}n+1}}(R_{\rm col,n})\sim (\sigma_{\rm
0,sh}R_{\rm col,n}/R_{\rm 0,sh})^{1/3}\sim \sigma_{\rm 0,sh}(R_{\rm
col,n}/R_{\rm c,sh})^{1/3}\sim\Gamma_{\rm 3B}(\sigma_{\rm
0,sh}/\Gamma_{\rm 3B})^{1/3}(n/N)^{2/(12-3k)}$, and reaches a
magnetization $\mean{\sigma_{{\rm sh,}n+1}}(R_{\rm col,n})\sim (R_{\rm
col,n}/R_{\rm c,sh})^{-1/3}\sim (\sigma_{\rm 0,sh}/\Gamma_{\rm
3B})^{2/3}(n/N)^{-2/(12-3k)}$.

\noindent{\bf Case 3A:}
for $R_{\rm cr, sh} < R_{\rm c,sh} < R_{\rm dec}\sim R_{\rm
cr}\tilde{\Delta}^{1/(4-k)}$ (or $\Gamma_{\rm cr,sh} < \sigma_{\rm
0,sh} < \Gamma_{\rm 3A}$ where $\Gamma_{\rm 3A}\sim (R_{\rm
dec}/R_{\rm 0,sh})^{1/2}\sim \Gamma_{\rm
cr}N^{1/2}\tilde{\Delta}^{1/(8-2k)}$ is the maximal attainable Lorentz
factor in the shell with mild magnetization, $\sigma\lesssim 1$;
bottom panel of Fig.~\ref{fig:sub-shells_k<2}) the trailing sub-shells
eventually (for $n/N\gtrsim (\sigma_0/\Gamma_{\rm 3B})^{4-k}$) become
kinetically dominated at $R_{\rm c,sh}$. They then start coasting and
spreading radially, thus quickly colliding and merging at $R_{\rm
IS}\sim (\Delta_{\rm gap}/\Delta_{\rm 0,sh})R_{\rm c,sh}$ where
$\mean{\sigma}\sim \Delta_{\rm 0,sh}/\Delta_{\rm gap}\lesssim 1$,
allowing for efficient energy dissipation in the resulting modest
magnetization internal shocks. The resulting merged modestly
magnetized shell is then gradually decelerated by a relativistic
reverse shock, largely following the unmagnetized ``thick shell'' case
at $R > R_{\rm IS}$ (so that $R_{\rm dec}\sim R_{\rm
cr}\tilde{\Delta}^{1/(4-k)}$, similar to case 2). Since $R_{\rm c,sh}
< R_{\rm dec}$ corresponds to $\sigma_{\rm 0,sh}\lesssim\Gamma_{\rm
3A}$, this implies that in case 3A the Lorentz factor of the resulting
internal shocks can be as high as $\mean{\Gamma} \sim
\sigma_0\lesssim\Gamma_{\rm 3A} \sim\Gamma_{\rm
cr}N^{1/2}\tilde{\Delta}^{1/(8-2k)}$, or up to a factor of $\sim
N^{1/2}\tilde{\Delta}^{1/(8-2k)}\gg 1$ larger than $\Gamma_{\rm cr}$.

\noindent{\bf Case 3B:}
for $R_{\rm c,sh} > R_{\rm dec}\sim R_{\rm
cr}\tilde{\Delta}^{2/(4-k)}$ (or $\sigma_{\rm 0,sh} > \Gamma_{\rm
3B}\sim \Gamma_{\rm cr}N^{1/2}\tilde{\Delta}^{1/(4-k)}$) even the last
sub-shells are still highly magnetized and do not spread radially
appreciably by the time they collide and merge with the preceding
sub-shell (as the latter is decelerated by the external medium or the
preceding sub-shell and spreads radially just before colliding with
the subsequent sub-shell).  Thus, subsequent collisions with later
ejected sub-shells proceed at the back end of the growing, highly
magnetized, quasi-uniform, merged shell behind the CD, whose Lorentz
factor evolves (on average) as $\Gamma_{\rm merged}\sim (L_{\rm
merged}/Ac^3)^{1/4}R^{(k-2)/4}$, until the deceleration
radius\footnote{In case 3B $R_{\rm dec}$ is slightly larger than in
cases 2 or 3A because the sub-shells remain highly magnetized near
$R_{\rm dec} \propto L_{\rm merged}(R_{\rm dec})^{-1/(4-k)}$, so that
the energy of each sub-shell decreases by a factor of
$\sim\tilde{\Delta}$ while its width grows by a similar factor,
resulting in $L_{\rm merged} \sim L/\tilde{\Delta}^2$ and $R_{\rm
dec}\sim R_{\rm cr}\tilde{\Delta}^{2/(4-k)}$. In case 2 (or 3A), all
(or late) sub-shells become kinetically dominated and their energy
remains constant until they are decelerated by the reverse shock,
while their width increases by a factor of $\tilde{\Delta}$ well
before $R_{\rm dec}$, so that $L_{\rm merged} \sim L/\tilde{\Delta}$
and $R_{\rm dec}\sim R_{\rm cr}\tilde{\Delta}^{1/(4-k)}$.}, $R_{\rm
dec}\sim R_{\rm cr}\tilde{\Delta}^{2/(4-k)}$. The last collision
occurs at $R_{{\rm col},N}\sim R_{\rm dec}$ and the last shell reaches
the maximal Lorentz factor of $\mean{\Gamma_{{\rm sh,}N}} \sim
\Gamma_{\rm 3B}(\sigma_{\rm 0,sh}/\Gamma_{\rm 3B})^{1/3}$, which is
larger than $\Gamma_{\rm 3B}$ by a factor of $(\sigma_{\rm
0,sh}/\Gamma_{\rm 3B})^{1/3} > 1$.  Since $R_{{\rm col},n}\sim R_{\rm
dec}(n/N)^{2/(4-k)}$, most of the collisions occur rather close to
$R_{\rm dec}$. Hence, case 3B largely follows the highly-magnetized
``thick shell'' case of regime II, with $L\to L_{\rm merged}\sim
L/\tilde{\Delta}^2$, where $\mean{\Gamma}(R<R_{\rm dec})\sim
(\sigma_{\rm 0,sh}R/R_{\rm 0,sh})^{1/3}$ is dominated by the almost
freely expanding and accelerating shells at the back of the flow. Near
$R_{\rm dec}$, essentially within a single dynamical time, all of the
remaining freely expanding shells collide and merge, so that
$\mean{\Gamma}$ moves from being dominated by those shells with a
typical Lorentz factor $\mean{\Gamma}(R_{{\rm dec,}-})\sim\Gamma_{\rm
3B}(\sigma_0/\Gamma_{\rm 3B})^{1/3}$ and magnetization
$\mean{\sigma}(R_{{\rm dec},}-)\sim\sigma_0/\mean{\Gamma}(R_{{\rm
dec,}-})\sim (\sigma_0/\Gamma_{\rm 3B})^{2/3}$ to being dominated by
the unmagnetized shocked external medium with $\mean{\Gamma}(R_{{\rm
dec,}+})\sim \Gamma_{\rm cr}\tilde{\Delta}^{-(3-k)/(4-k)}$, where near
$R_{\rm dec}$ a significant fraction of the total energy also resides
in the merged highly magnetized outflow shell, which has a typical
Lorentz factor $\mean{\Gamma}(R_{{\rm dec,}+})$ and magnetization
$\mean{\sigma}(R_{{\rm
dec},}+)\sim\sigma_0/\tilde{\Delta}\mean{\Gamma}(R_{{\rm
dec,}+})\sim\tilde{\Delta}^{-1/(4-k)}\sigma_0/\Gamma_{\rm cr}\gg 1$.

\subsection{Stronger external density stratification: $2 < k < 3$}
\label{sec:2<k<3}

\begin{figure}
\centerline{\includegraphics[height=6.5cm]{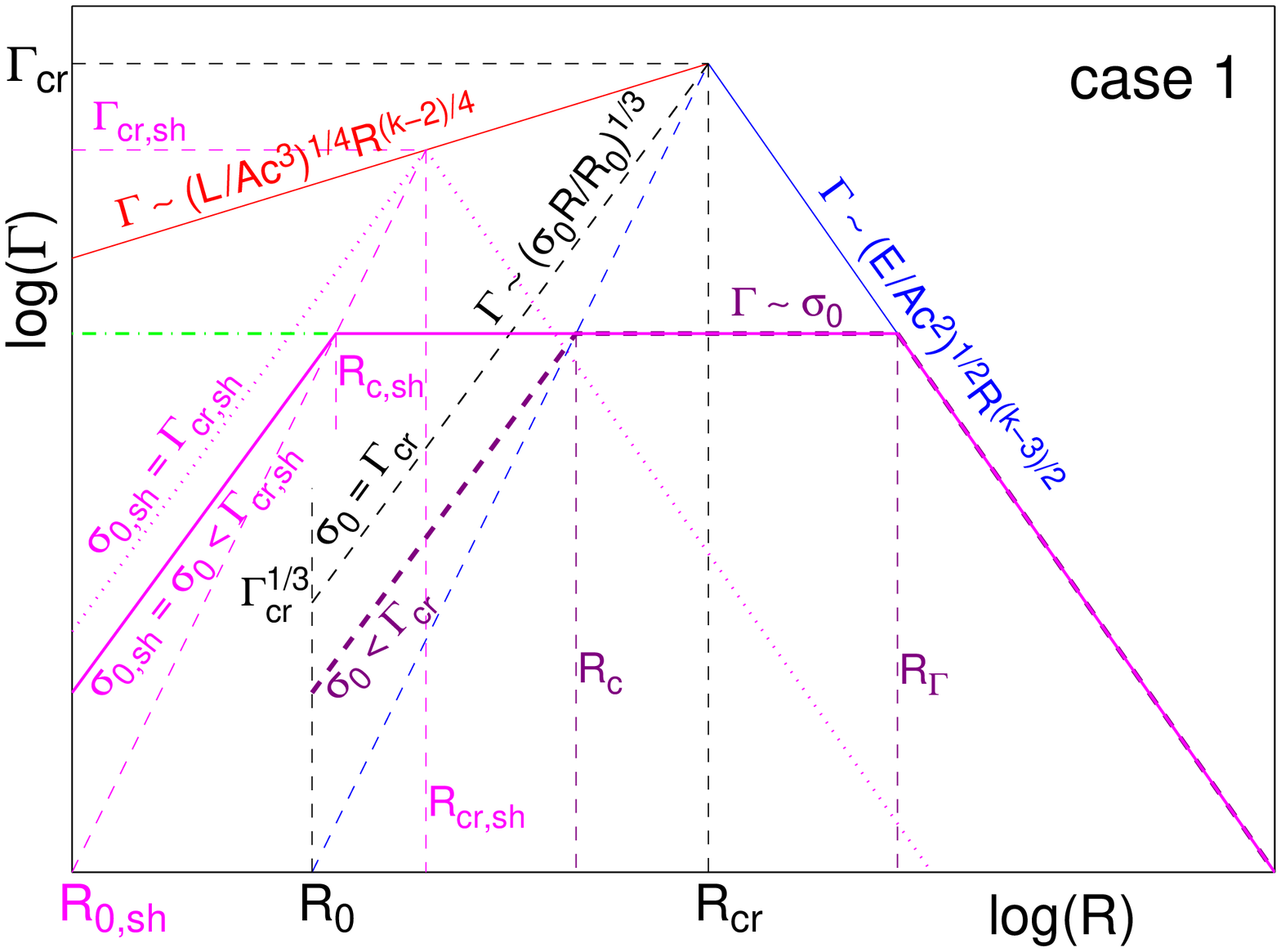}}
\vspace{0.5cm}
\centerline{\includegraphics[height=6.5cm]{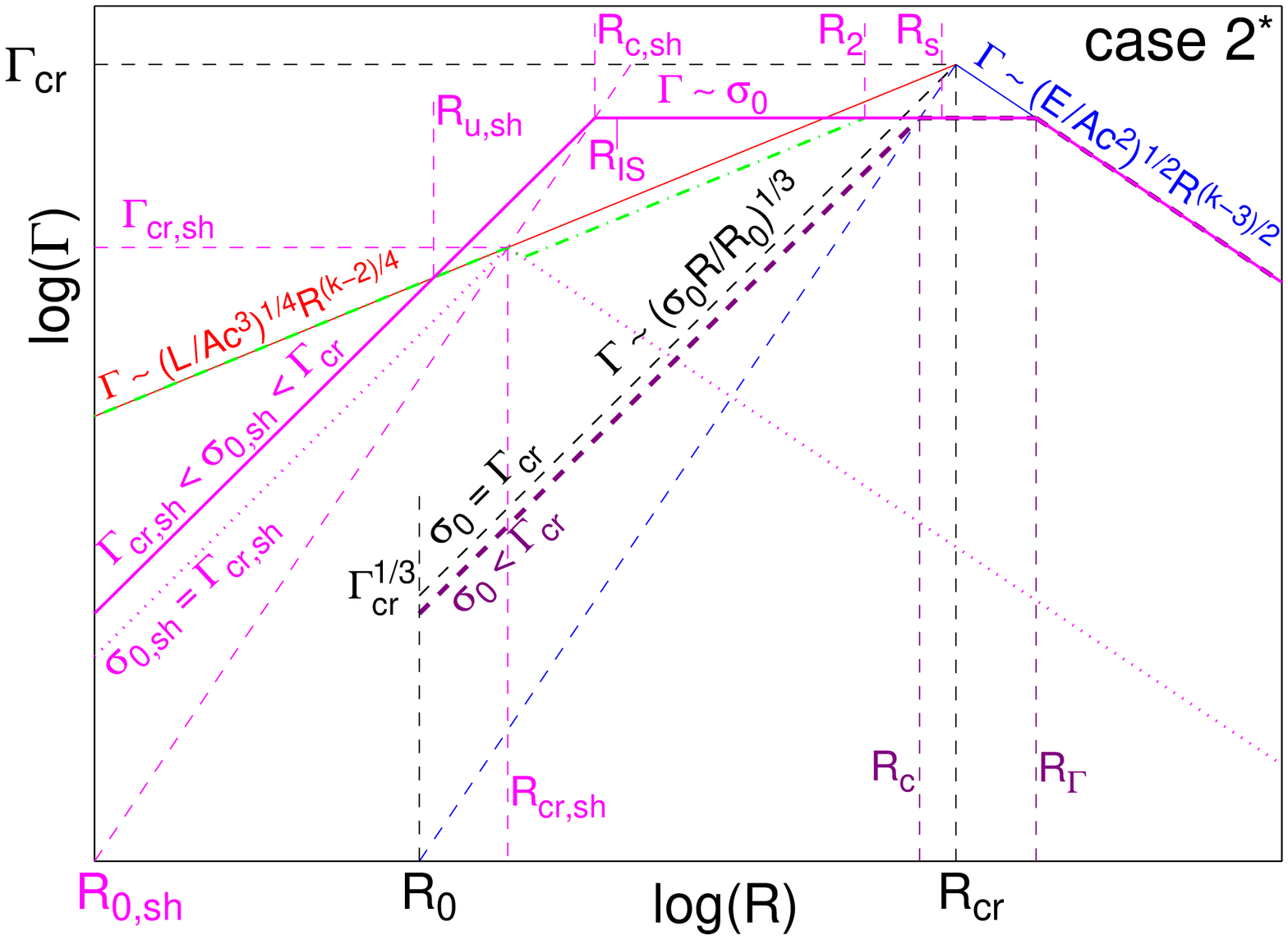}}
\vspace{0.5cm}
\centerline{\includegraphics[height=6.5cm]{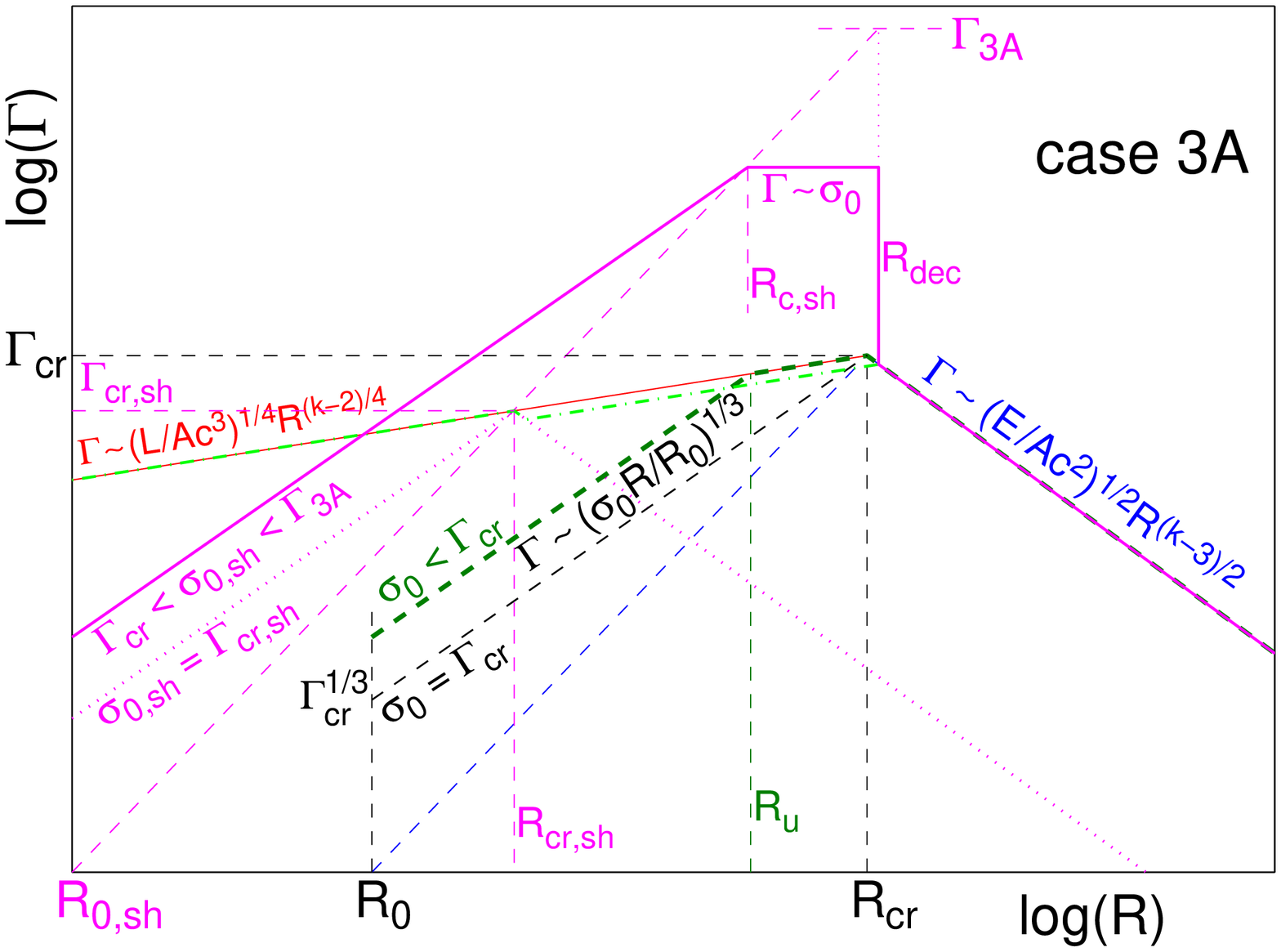}}
\caption{
The same as Fig.~\ref{fig:sub-shells_k<2} but for $2<k<3$.}
\label{fig:sub-shells_2<k<3}
\end{figure}

\noindent{\bf Case 1}
now corresponds to $\sigma_{\rm 0,sh} = \sigma_0 < \Gamma_{\rm cr,sh}$
(i.e. $\sigma_0/\Gamma_{\rm cr} \sim (R_c/R_{\rm cr})^{1/2} <
N^{(2-k)/(8-2k)}$) or $R_{\rm c,sh} < R_{\rm cr,sh}$ (i.e. $R_{\rm
c,sh}/R_{\rm cr} < N^{-2/(4-k)}$), but otherwise the behavior of the
system in this regime is very similar to that in case 1 for $k<2$ (see
top panel of Fig.~\ref{fig:sub-shells_2<k<3}).

\noindent{\bf Case 2$^*$:} what used to be case 2 for $k<2$
now corresponds to $\Gamma_{\rm cr,sh} < \sigma_0=\sigma_{\rm 0,sh} <
\Gamma_{\rm cr}$ (i.e. $N^{(2-k)/(8-2k)}<\sigma_0/\Gamma_{\rm cr}<1$) or
$R_{\rm cr,sh} < R_{\rm c,sh} < R_{\rm 0,sh}\Gamma_{\rm cr}^2$
(i.e. $N^{-2/(4-k)} < R_{\rm c,sh}/R_{\rm cr} < N^{-1}$), and is
called case $2^*$ since its properties are somewhat different (see
middle panel of Fig.~\ref{fig:sub-shells_2<k<3}). The first sub-shell
is in regime II, i.e. it starts to be significantly affected by the
external medium (accelerates more slowly, as $\mean{\Gamma_{\rm
sh}}\propto R^{(k-2)/4}$ instead of $R^{1/3}$) at $R_{u,{\rm sh}} \sim
R_u N^{-4/(10-3k)}$. The first few subsequent sub-shells would still
collide with the back of the growing highly magnetized merged shell
behind the CD. However, once the radius of this merged shell (or of
the CD) exceeds $R_{\rm c,sh}$ the sub-shells first become kinetically
dominated at $R_{\rm c,sh}$, start coasting at $\mean{\Gamma_{\rm
sh}}\sim\sigma_{\rm 0,sh}$, expand radially, and collide with each
other (at $R_{\rm IS}\sim (\Delta_{\rm gap}/\Delta_{\rm 0,sh})R_{\rm
c,sh}$) with a moderate magnetization ($\mean{\sigma}(R_{\rm
IS})\sim\Delta_{\rm 0,sh}/\Delta_{\rm gap}\lesssim 1$) before
effectively colliding with (or being decelerated by) the back end of
the merged shell. Therefore, once they start being decelerated by the
merged shell behind the CD (which has $\Gamma \sim \Gamma_{\rm CD}$)
it occurs in the form of a highly relativistic reverse shock, as long
as $\Gamma_{\rm CD}\ll\sigma_{\rm 0,sh}$. However, this shock becomes
Newtonian and weak at $R_2\sim\tilde{\Delta}^{-1/(2-k)}R_1$ (where
$\Gamma_{\rm CD}\sim\sigma_{\rm 0,sh}$), and from that point on the
merged mildly magnetized shell essentially coasts at $\Gamma \sim
\sigma_{\rm 0,sh} = \sigma_0$. The resulting merged shell of width
$\sim\Delta_{\rm tot}\approx \Delta_0\tilde{\Delta} \sim
\Delta_0(\Delta_{\rm gap}/\Delta_{\rm 0,sh})$ starts spreading
radially only at $R_s\sim \Delta_{\rm tot}\sigma_{\rm 0,sh}^2\sim
(\Delta_{\rm gap}/\Delta_{\rm 0,sh})R_c\sim NR_{\rm IS}$ where its
mean magnetization is still similar to that near the internal shock
radius, $\mean{\sigma}(R_s) \sim \mean{\sigma}(R_{\rm
IS})\sim\Delta_{\rm 0,sh}/\Delta_{\rm gap}\lesssim 1$, and evolves
similarly to case 1. At $R>R_s$ it starts evolving as $\Delta \sim
(R/R_s)\Delta_{\rm tot} \sim (R/R_c)\Delta_0$ and $\mean{\sigma} \sim
(R_s/R)\mean{\sigma}(R_s)\sim R_c/R$, similar to the original single
wide shell, thus following the unmagnetized ``thin shell'' case. One
possible difference is that the early short-lived phase of a strong
relativistic reverse shock might result in an observable early and
short-lived spike in the reverse shock emission, on a timescale of
$\sim (R_2/R_\Gamma)T_{\rm dec}\sim (\Gamma_{\rm
cr}/\sigma_0)^{(8-2k)/[(3-k)(2-k)]}T_{\rm dec}$ that is precedes the
main reverse shock emission peak (which peaks on a larger timescale of
$T_{\rm dec}$). For $k<3$ the shell is eventually decelerated at
$R_{\rm dec} \sim R_\Gamma$ by the reverse shock, which becomes mildly
relativistic by that radius, and then the flow approaches the
\citet{BM76} self-similar solution.

\noindent{\bf Case 3} now corresponds to
$\sigma_{\rm 0,sh} = \sigma_0 > \Gamma_{\rm cr}$
(i.e. $\sigma_0/\Gamma_{\rm cr} \sim (R_c/R_{\rm cr})^{1/2} > 1$) or
$R_{\rm c,sh} > R_{\rm 0,sh}\Gamma_{\rm cr}^2\sim R_{\rm cr}/N$.  {\bf
Case 3A} corresponds to $R_{\rm 0,sh}\Gamma_{\rm cr}^2 < R_{\rm c,sh}
< R_{\rm dec}$ (i.e. $N^{-1} < R_{\rm c,sh}/R_{\rm cr} <
\tilde{\Delta}^{1/(4-k)}$) or $\Gamma_{\rm cr} < \sigma_0 <
\Gamma_{\rm 3A} \sim
\Gamma_{\rm cr}N^{1/2}\tilde{\Delta}^{1/(8-2k)}$ (see
bottom panel of Fig.~\ref{fig:sub-shells_2<k<3}), while {\bf Case 3B:}
corresponds to $R_{\rm c,sh} > R_{\rm dec}$ (i.e. $R_{\rm c,sh}/R_{\rm cr} >
\tilde{\Delta}^{2/(4-k)}$) or $\sigma_0 > \Gamma_{\rm
3B}$. Other than that, cases 3A and 3B behave very similarly to $k<2$
(described above).

\subsection{A wind-like external density profile: $k = 2$}
\label{sec:k=2}

For $k= 2$ we have $\Gamma_{\rm cr,sh} = \Gamma_{\rm cr}$ and $R_{\rm
cr,sh} = R_{\rm 0,sh}\Gamma_{\rm cr}^2$, so that there is no case 2 or
$2^*$.  {\bf Case 1} corresponds to $\sigma_{\rm 0,sh} = \sigma_0 <
\Gamma_{\rm cr} = \Gamma_{\rm cr,sh}$ (or $R_{\rm c,sh}< R_{\rm cr,sh}
= R_{\rm 0,sh}\Gamma_{\rm cr}^2$). {\bf Case 3A} corresponds to
$\Gamma_{\rm cr,sh} = \Gamma_{\rm cr} <\sigma_{\rm 0,sh} < \Gamma_{\rm
3A}$) (or $R_{\rm cr, sh} = R_{\rm 0,sh}\Gamma_{\rm cr}^2 < R_{\rm
c,sh} < R_{\rm dec}$). {\bf Case 3B } corresponds to $\sigma_{\rm
0,sh} > \Gamma_{\rm 3B}$ (or $R_{\rm c,sh} > R_{\rm dec}$).

\section{Varying the initial magnetization $\sigma_{\rm 0,sh}$ between different sub-shells}
\label{sec:vary-sigma}

It is reasonable to expect that the magnetization of different
sub-shells might differ, at least by factors of order unity. Here we
consider the effects of such a variation, while keeping the other
sub-shell parameters fixed (namely $\Delta_{\rm 0,sh}\approx R_{\rm
0,sh}$, $\Delta_{\rm gap}$, $L_{\rm sh} \approx L$).

First, let us examine whether sub-shells might collide during the
acceleration stage. Consider two sub-shells ejected with a time
difference $t_{\rm gap}\approx\Delta_{\rm gap}/c$, the first with
$\sigma_{\rm 0,sh} = \sigma_{0,1}$ and the second with $\sigma_{\rm
0,sh} =\sigma_{0,2} > \sigma_{0,1}$. Each sub-shell initially
accelerates as $\Gamma_i \sim (\sigma_{0,i}R/R_{\rm 0,sh})^{1/3}$ and
its width remains almost constant in the lab frame up to its coasting
radius, $R_{{\rm c,}i}\sim R_{\rm 0,sh}\sigma_{0,i}^2$, so that both
sub-shells accelerate at $R < R_{\rm c,1} < R_{\rm c,2}$. Hence, when
making the simplifying assumption of uniform sub-shells, the
separation between them evolves as
\begin{eqnarray}\nonumber
l(R<R_{\rm c,1}) 
&\approx& \Delta_{\rm gap} - 
\int_{R_{\rm 0,sh}}^{R}\frac{dR}{2}\left(\frac{1}{\Gamma_1^2}-\frac{1}{\Gamma_2^2}\right)
\\
&\approx& \Delta_{\rm gap} - \frac{3\Delta_{\rm 0,sh}}{2\sigma_{0,1}^{2/3}}
\left[1-\fracb{\sigma_{0,1}}{\sigma_{0,2}}^{2/3}\right]
\left[\fracb{R}{R_{\rm 0,sh}}^{1/3}-1\right]\ .
\end{eqnarray}
The two sub-shells would effectively collide when $l(R) = 0$,
corresponding to a collision radius $R_{\rm col}\gg R_{\rm 0,sh}$,
which for $R_{\rm col}\leq R_{\rm c,1}$ is given by
\begin{equation}
R_{\rm col}(R_{\rm col}\leq R_{\rm c,1}) \approx R_{\rm
c,1}\fracb{2\Delta_{\rm gap}}{3\Delta_{\rm
0,sh}}^3\left[1-\fracb{\sigma_{0,1}}{\sigma_{0,2}}^{2/3}\right]^{-3}\ . 
\end{equation}
Thus, for $\Delta_{\rm gap} > 1.5[1-(\sigma_{0,1}/\sigma_{0,2})^{2/3}]
\Delta_{\rm 0,sh}$ such a collision would occur at $R_{\rm col} >
R_{\rm c,1}$, after the first (and slower) shell reaches its coasting
radius and becomes kinetically dominated. This condition always holds
for $\Delta_{\rm gap} > 1.5\Delta_{\rm 0,sh}$, regardless of the ratio
of the initial sub-sell magnetizations, $\sigma_{0,2}/\sigma_{0,1} >
1$, and reasonably low values of this ratio relax the condition on
$\Delta_{\rm gap}/\Delta_{\rm 0,sh}$ (e.g. it becomes $\Delta_{\rm
gap}/\Delta_{\rm 0,sh} > 1$ for $\sigma_{0,2}/\sigma_{0,1} =
3^{3/2}\approx 5.2$).

For $\Delta_{\rm gap}/\Delta_{\rm 0,sh} < 1$ one cannot neglect the
radial spreading of the first sub-shell before its coasting radius,
$R_{\rm c,1}$, since it would result in a collision at $R_{\rm IS}
\sim (\Delta_{\rm gap}/\Delta_{\rm 0,sh})^3 R_{\rm c,1} < R_{\rm
c,1}$. For $\Delta_{\rm gap}/\Delta_{\rm 0,sh} > 1$ the effect of this
radial spreading on the decrease in the separation between the two
sub-shells becomes comparable to or larger than that of the difference
in their typical Lorentz factors at $R
\gtrsim R_{\rm c,1}$. Thus, when accounting for both of these effects,
for $\Delta_{\rm gap}/\Delta_{\rm 0,sh}\gtrsim 1\,$--$\,1.5$ the two
sub-shells effectively collide near $R_{\rm IS}\sim (\Delta_{\rm
gap}/\Delta_{\rm 0,sh})R_{\rm c,1} > R_{\rm c,1}$. For a large
contrast in the initial magnetizations, $\sigma_{0,2}/\sigma_{0,1} >
(\Delta_{\rm gap}/\Delta_{\rm 0,sh})^{1/2}$, the second sub-shell
would still be highly magnetized at the collision radius,
$\mean{\sigma_{\rm sh,2}}(R_{\rm IS})\sim (R_{\rm c,2}/R_{\rm
IS})^{1/3}\sim (\Delta_{\rm gap}/\Delta_{\rm
0,sh})^{-1/3}(\sigma_{0,2}/\sigma_{0,1})^{2/3} > 1$, while for a mild
initial magnetization contrast, $\sigma_{0,2}/\sigma_{0,1} \lesssim
(\Delta_{\rm gap}/\Delta_{\rm 0,sh})^{1/2}$, it would be mildly
magnetized, $\mean{\sigma_{\rm sh,2}}(R_{\rm IS}) \lesssim 1$. The
first sub-shell would always be mildly magnetized in this regime,
$\mean{\sigma_{\rm sh,1}}(R_{\rm IS}) \sim \Delta_{\rm
0,sh}/\Delta_{\rm gap} < 1$. This would allow for reasonably efficient
dissipation in the resulting internal shocks.

If the trailing sub-shell has a lower initial magnetization than the
leading one ($\sigma_{0,2}<\sigma_{0,1}$) then the (effective)
separation between the sub-shells initially increases during the
acceleration stage. Once the trailing sub-shell becomes kinetically
dominated and starts coasting at $R_{\rm c,2} < R_{\rm c,1}$ it starts
spreading radially, but even if its head moves at very close to the
speed of light it would effectively collide with the bulk of the first
sub-shell only after the latter starts spreading appreciably. Such a
spreading of the first sub-shell can occur either if it reaches its
coasting radius ($R_{\rm c,1}$) and becomes kinetically dominated (in
which case both sub-shells would have a mild or low magnetization when
they collide), or alternatively if it is still highly magnetized but
decelerates as it transfers a good part of its energy to the sub-shell
in front of it (or the shocked external medium across the CD) through
$PdV$ work.

Altogether, a reasonable spread in the initial magnetization of the
sub-shells, of $\delta\sigma_{\rm 0,sh} \sim \sigma_{\rm 0,sh}$, would
not have a very large effect on the overall dynamics or the efficiency
of the resulting internal shocks. Even a very high contrast, of
$\delta\sigma_{\rm 0,sh} \gg \sigma_{\rm 0,sh}$, is expected to
typically affect the overall efficiency of the internal shocks only by
a factor of order unity (since at least one of the sub-shells in each
collision is expected to have mild or low magnetization). Moreover, it
is not even clear whether the overall efficiency would actually be
decreased or increased, since while some of the collisions might occur
where one of the colliding sub-shells has a higher magnetization
(compared to the case $\delta\sigma_{\rm 0,sh} \ll \sigma_{\rm
0,sh}$), the relative Lorentz factor of the colliding shells can be
higher, thus increasing the the efficiency of the dissipation in the
low magnetization colliding sub-shell. Similarly, order unity
variations in $\Delta_{\rm 0,sh}$, $\Delta_{\rm gap}$, or $L_{\rm sh}
= E_{\rm sh}c/\Delta_{\rm 0,sh}$ are not expected to have a very large
effect on the overall dynamics or on the efficiency of the resulting
internal shocks or reverse shock.

As for the reverse shock, if some of the sub-shells remain highly
magnetized within the merged shell that accumulates behind the CD,
then this could suppress the reverse shock as it passes these
sub-shells, thus effectively causing its emission to turn on and off
as it passes regions (corresponding to different original sub-shells)
of low and high magnetization. Such a variable reverse shock emission
might have been observed in some cases [e.g., GRB~080319B
\citep{Racusin08}, GRB~070419A \citep{Melandri09}, GRB~110205A
\citep{Cucchiara11}].

\section{Summary}
\label{sec:sum}

To summarize, when the outflow consists of a large number of well
separated sub-shells there are three main regimes. One of them -- the
highly magnetized ``thick shell'' (case 3B; $R_{\rm c,sh} > R_{\rm
dec} \sim R_{\rm cr}\tilde{\Delta}^{2/(4-k)}$ or $R_{\rm c,sh}/R_{\rm
cr} > \tilde{\Delta}^{2/(4-k)}$) is expected to result in a severe
suppression of the reverse shock and its associated
emission. Nevertheless, the thick shell nature of this regime can be
deduced from the fact that the onset time of the afterglow emission,
$T_{\rm dec}$, is similar to the duration of the prompt GRB
emission ($T_{\rm dec}\sim T_{\rm GRB}$).  Therefore, this leaves
two regimes for which a reasonably bright reverse shock emission may
occur.
In the low magnetization ``thick shell'' case (i.e. cases 2 or 3A;
$N^{-1} < R_{\rm c,sh}/R_{\rm cr} < \tilde{\Delta}^{1/(4-k)}$), which
was described in this work, the mean magnetization of the merged shell
at the time when its bulk is crossed by the relativistic reverse shock
(near $R_{\rm dec}\sim\tilde{\Delta}^{1/(4-k)}R_{\rm cr}$) is
$\mean{\sigma}(R_{\rm dec})\sim \mean{\sigma}(R_{\rm
IS})\sim\Delta_{\rm 0,sh}/\Delta_{\rm gap}\lesssim 1$, i.e. it is
expected to be less than unity but not by a large factor
($0.1-0.3\lesssim\mean{\sigma}(R_{\rm dec})\lesssim 1$). Thus, very
low values of $\mean{\sigma}$ are not expected in this regime, whose
main observational signature is that the afterglow emission and the
reverse shock emission both peak on a timescale similar to the
duration of the prompt GRB emission ($T_{\rm dec}\sim T_{\rm GRB}\sim
(1+z)\Delta_{\rm tot}/c$). The expectations of this regime (both in
terms of $T_{\rm dec}\sim T_{\rm GRB}$ and the value of
$\mean{\sigma}(R_{\rm dec})$) appear consistent with the bright prompt
optical emission from GRB~990123, which had been attributed to the
reverse shock~\citep{Akerlof99,SP99,Fan02,Zhang03,NP05}.
In the low magnetization ``thin shell'' case (regime I; cases 1 or
$2^*$; $R_{\rm c,sh}/R_{\rm cr} < N^{-1}$), the deceleration time
$T_{\rm dec}$ that corresponds to the duration of the peak reverse
shock and afterglow emission components is expected to be larger than
the prompt GRB duration, $T_{\rm dec} \gg T_{\rm GRB}$, and the
magnetization at the radius where most of the energy is dissipated in
the reverse shock (near $R_{\rm dec} \sim R_\Gamma$) is expected to be
$\mean{\sigma}(R_{\rm dec})\sim (\sigma_0/\Gamma_{\rm
cr})^{2(4-k)/(3-k)} \sim R_c/R_{\rm dec} \sim T_{\rm GRB}/(T_{\rm
dec}\tilde{\Delta})\ll 1$, i.e. a factor of $\sim T_{\rm dec}/T_{\rm
GRB}\gg 1$ smaller than in the low magnetization ``thick shell''
case. Thus, a clear prediction of this model is a positive linear
correlation between $T_{\rm GRB}/T_{\rm dec}$ and
$\mean{\sigma}(R_{\rm dec})$ when there is a bright reverse shock
emission (i.e. when $\mean{\sigma}(R_{\rm dec})\lesssim\Delta_{\rm
0,sh}/\Delta_{\rm gap}\lesssim 1$).

Moreover, this model also has predictions for the internal shocks,
which could be tested against observations if the prompt GRB emission
is indeed from such internal shocks. The internal shocks radius is given
by
\begin{equation}
R_{\rm IS} \sim \frac{\Delta_{\rm gap}}{\Delta_{\rm 0,sh}}R_{\rm c,sh}
\sim 10^{14}\,\frac{\Delta_{\rm gap}}{\Delta_{\rm 0,sh}}\zeta^{-1}
\fracb{\sigma_{\rm 0,sh}}{10^{2.5}}^2\fracb{T_{\rm var,obs}}{0.1\;{\rm s}}\;{\rm cm}\ ,
\end{equation}
which satisfies the usual relation, $R_{\rm IS} \sim \Gamma^2(R_{\rm
IS})\,c\,T_{\rm var,obs}$, where $T_{\rm var,obs}$ is the observed
variability time in the prompt GRB lightcurve.  For reference,
\begin{equation}\label{R_cr_acc}
R_{\rm cr} \approx 
\left\{\matrix{9.3\times 10^{16}a^{1/4}\zeta^{-1/4}\tilde{\Delta}^{-1/4}n_0^{-1/4}
E_{53}^{1/4}T_{30}^{1/4}\;{\rm cm} & (k = 0)\ , \cr\cr
5.3\times 10^{15}a^{1/2}\zeta^{-1/2}\tilde{\Delta}^{-1/2}A_*^{-1/2}
E_{53}^{1/2}T_{30}^{1/2}\;{\rm cm} & (k = 2)\ ,}\right.
\end{equation}
(see Eqs.~[23] and [45] of paper II for the definition of $a$ and the
derivation of $R_{\rm cr}$) and
\begin{equation}\label{R_cr_IS}
\frac{R_{\rm c,sh}}{R_{\rm cr}} \approx \left\{\matrix{
10^{-2.5}\zeta^{-3/4}\tilde{\Delta}^{-3/4}
a^{-1/4}n_0^{1/4}E_{53}^{-1/4}T_{30}^{3/4}\fracb{N}{100}^{-1}
\fracb{\sigma_{\rm 0,sh}}{10^{2.5}}^{2} & (k = 0)\ , \cr\cr
0.056\zeta^{-1/2}\tilde{\Delta}^{-1/2}
a^{-1/2}A_*^{1/2}E_{53}^{-1/2}T_{30}^{1/2}\fracb{N}{100}^{-1}
\fracb{\sigma_{\rm 0,sh}}{10^{2.5}}^{2} & (k = 2)\ .}\right.
\end{equation}
This shows that the relevant regimes correspond to reasonable model
parameters, and could potentially occur in different GRBs. The most
uncertain parameter is the initial magnetization, $\sigma_{\rm 0,sh}$
or $\sigma_0$, whose value can be estimated in the low magnetization
regimes, where the outflow becomes kinetically dominated (with
$\Gamma\sim\sigma_0$; lower limits on $\Gamma$ also serve as lower
limits on $\sigma_0$ in this model, without requiring kinetic
dominance). Current constraints from pair opacity in the prompt
emission and from the onset of the afterglow suggest
$10^2\lesssim\sigma_0\lesssim 10^3$.
In case 3B, which corresponds to $R_{\rm c,sh}/R_{\rm cr}
>\tilde{\Delta}^{2/(4-k)} > 1$, all of the sub-shells collide while
they are still highly magnetized, which suppresses the internal shocks
and their associated emission, making them unlikely to power the
prompt GRB emission (which in this case might be alternatively powered
by magnetic reconnection events in the highly magnetized outflow). In
all other cases the internal shocks occur at mild or low
magnetization, allowing them to be reasonably efficient and
potentially power the prompt GRB emission.

\section{Discussion}
\label{sec:dis}

The effects of sub-shells in an impulsive, initially highly magnetized
relativistic outflow have been studied, and compared to the case of a
single wide shell. It has been shown that if a single wide uniform
outflow shell is divided into a large number of sub-shells with a
reasonable initial contrast and spacing between them ($\Delta_{\rm
gap}\gtrsim\Delta_{\rm 0,sh}$) then it could reach a significantly
higher Lorentz factor.\footnote{\citet{Levinson10} has argued that for
$\Delta_{\rm gap}\sim\Delta_{\rm 0,sh}$ the sub-shells would
effectively collide an merge well before their coasting radius (by a
factor of $\sim\sigma_{\rm 0,sh}^{2/3}\gg 1$), while they are still
highly magnetized ($\mean{\sigma_{\rm sh}}\sim\sigma_0^{2/9}\gg 1$),
and would thus have a very small effect on the outflow, and in
particular would not help to increase its maximal Lorentz
factor. However, this conclusion is wrong and arises due to an error
in his Eq.~(29), which results in an incorrect expression for the
collision time or radius.}  The leading sub-shells effectively clear
the way for the subsequent sub-shells, allowing them to accelerate for
a longer time without feeling the effects of the external medium
(almost as if into vacuum), thus enabling them to reach a higher
Lorentz factor. Moreover, internal shocks arise from collisions
between different sub-shells, which naturally occur at relatively high
Lorentz factors and at low magnetizations that are vital in order to
have a reasonable energy dissipation efficiency in the internal
shocks.

A sufficiently high Lorentz factor is needed to overcome the
compactness problem and avoid excessive pair production within the
source~\citep{KP91,Fenimore93,WL95,BH97,LS01}.  It has been recently
argued~\citep{Levinson10} that the interaction with the external
medium might not enable an impulsive highly magnetized outflow in GRBs
to accelerate up to sufficiently high Lorentz factors, and in
particular that its maximal achievable Lorentz factor is largely
limited to $\Gamma \lesssim \Gamma_{\rm cr}$. This would pose a
particularly severe problem for a stellar wind-like external medium
($k=2$) for which typically $\Gamma_{\rm cr}\lesssim 10^2$ (see
Eq.~[\ref{eq:Gamma_cr}]). Recent high-energy observations by the Fermi
Large Area Telescope (LAT) have set a lower limit of $\Gamma\gtrsim
10^3$ for the emitting region in a number of GRBs with a bright
high-energy emission~\citep{GRB080916C,GRB090902B,GRB090510} using a
simplified one-zone model. However, a more detailed and realistic
treatment shows that the limit is lower by a factor of $\sim
3$~\citep{Granot08,GRB090926A}, which would correspond to
$\Gamma\gtrsim 10^{2.5}$ for the brightest Fermi LAT GRBs \citep[see
also][]{HDMV11}. This might nevertheless still pose a problem for a
single highly magnetized shell in a stellar-wind environment. The
present work, however, shows that if it is divided into a large number
of sub-shells, then its Lorentz factor $\Gamma$ could exceed
$\Gamma_{\rm cr}$ by up to a factor of $\sim\Gamma_{\rm
3A}/\Gamma_{\rm cr}\sim N^{1/2}\tilde{\Delta}^{1/(8-2k)} \gg
1$. Moreover, most of the dissipation in internal shocks is expected
to occur near the maximal Lorentz factor attained by the outflow. This
would greatly help satisfy the lower limits on $\Gamma$ from
compactness arguments, or from the onset of the afterglow
emission~\citep[usually around a few
hundred;][]{SP99,NP05,Molinari07,ZP10,Gruber11}, also for a stellar
wind environment.

It has been found (in papers I and II) that for a single shell there
are two main dynamical regimes: the low magnetization ``thin shell''
(regime I, where the shell becomes kinetically dominated, coasts and
spreads radially, and a reverse shock develops that becomes mildly
relativistic near the deceleration radius, $R_{\rm dec}\sim R_\Gamma$,
and whose emission peeks at $T_{\rm dec}
\gg T_{\rm GRB}$), and the high magnetization ``thick shell'' (regime II
or III, where the shell remains highly magnetized without reaching a
coasting stage or spreading radially, and the the reverse shock is
suppressed along with its associated emission). A high magnetization
``thin shell'' or a low magnetization ``thick shell'' are not possible
for a single initially highly magnetized shell. In this work, it has
been shown that a low magnetization ``thick shell'' regime becomes
possible (and occurs in cases 2 or 3A) if such an initially highly
magnetized shell is divided into a large number of sub-shells with
reasonable initial separations ($\Delta_{\rm gap}\gtrsim\Delta_{\rm
0,sh}$). This would allow a relativistic reverse shock with bright
emission on a timescale comparable to that of the prompt GRB emission
($T_{\rm dec}\sim T_{\rm GRB}$).  Moreover, if there are large
variations in the magnetization between different sub-shells, this
might result in alternating regions of high and low magnetization
through which the reverse shock passes, causing it and its associated
emission to be alternately suppressed and revived, resulting in a
variable reverse shock emission. Such a reverse shock emission may
bear some temporal correlation to the prompt emission from the
internal shocks~\citep[somewhat analogous to the pure hydrodynamic
case;][]{NP04}, as both are affected by the magnetization of the
sub-shells, though some delay might be expected between corresponding
features in the internal shocks and in the reverse shock.

\acknowledgements
The author thanks A. Spitkovsky, Y.~E. Lyubarsky, T. Piran, A. Levinson and
S.~S. Komissarov for useful comments on the manuscript.  This research
was supported by the ERC advanced research grant ``GRBs''.

\newpage
\appendix

\section{Estimating the collision radius of sub-shells with the uniform region behind the CD}
\label{app:R_col}

I consider here sub-shells that are still highly magnetized and have
not spread radially significantly before colliding with the previous
sub-shell that is decelerated by its $PdV$ work on the shocked
external medium across the CD (or on the preceding sub-shell).  Let
us begin with the first such collision. The first sub-shell is
significantly decelerated by the external medium at $R_{\rm u,sh}\sim
R_u N^{-4/(10-3k)}$, at which stage its separation from the subsequent
shell is still close to its initial value, $\Delta_{\rm gap}$. One way
of estimating the collision radius is that the head of the second
sub-shell travels faster than the tail of the second sub-shell so that
the difference in their velocities in the lab frame is $\Delta\beta
\approx 1/2\Gamma_{\rm CD}^2 \propto R^{(2-k)/2}$ and their effective separation changes
with radius as
\begin{eqnarray}\nonumber
l(R>R_{\rm u,sh}) &\approx & \Delta_{\rm gap} - 
\int_{R_{\rm u,sh}}^{R}\frac{dR}{2\Gamma_{\rm CD}^2(R)}
= \Delta_{\rm gap} - \frac{R_{\rm u,sh}}{2\Gamma_{\rm CD}^2(R_{\rm u,sh})}
\int_{1}^{R/R_{\rm u,sh}}d\tilde{R}\tilde{R}^\frac{2-k}{2}
\\ \label{eq:R_col01}
&=& \Delta_{\rm gap} - \frac{R_{\rm u,sh}}{(4-k)\Gamma_{\rm CD}^2(R_{\rm u,sh})}
\left[\fracb{R}{R_{\rm u,sh}}^{(4-k)/2}-1\right]\ .
\end{eqnarray}
Since the sub-shell's Lorentz factor before it is significantly
decelerated is $\Gamma_{\rm sh}(R<R_{\rm u,sh}) \sim (\sigma_{\rm
0,sh}R/R_{\rm 0,sh})^{1/3}$, we have $R_{\rm u,sh}/\Gamma_{\rm
CD}^2(R_{\rm u,sh}) \sim \Delta_{\rm 0,sh}(R_{\rm
u,sh}/R_{\rm c,sh})^{1/3} \sim\Delta_{\rm 0,sh}(R_{\rm cr,sh}/R_{\rm
u,sh})^{(k-4)/2}$ or $\Delta_{\rm gap}\Gamma_{\rm CD}^2(R_{\rm
u,sh})/R_{\rm u,sh} \approx (\Delta_{\rm
gap}/\Delta_{\rm 0,sh})(R_{\rm cr,sh}/R_{\rm u,sh})^{(4-k)/2}$. The
collision occurs when the separation reaches zero, and in the
relevant regime the second term in the square brackets in
Eq.~(\ref{eq:R_col01}) can can neglected, giving a collision radius of
\begin{equation}
R_{\rm col,1} \approx \left[(4-k)\frac{\Delta_{\rm gap}}{\Delta_{\rm
0,sh}}\right]^{2/(4-k)}R_{\rm cr,sh} \sim \tilde{\Delta}^{2/(4-k)}R_{\rm cr,sh}\ .
\end{equation}
A comparable estimate of $R_{\rm col,1}$ is obtained when asking at
what radius the radial spreading of the first sub shell due to the
dispersion in its Lorentz factor ($\delta\Gamma_{\rm
sh}\sim\mean{\Gamma_{\rm sh}}$) and deceleration because of the work
it performs on the shocked external medium across the CD becomes large
enough to bridge the initial gap, $\Delta_{\rm gap}$, from the
subsequent sub-shell. This is since taking $l(R) = 0$ in
Eq.~(\ref{eq:R_col01}) is essentially equivalent to the requirement on
the spreading of the shell,
\begin{equation}
\Delta_{\rm gap} = \Delta_{\rm sh}(R_{\rm col,1})-\Delta_{\rm 0,sh} \sim
\int_{R_{\rm u,sh}}^{R_{\rm col,1}}\frac{dR}{2\Gamma_{\rm CD}^2(R)}\ .
\end{equation}

The above considerations can be readily generalized in order to
estimate when a point with an initial lag of $\Delta$ moving at
$\Gamma\gg\Gamma_{\rm CD}$ catches up with the CD. Simply replacing
$\Delta_{\rm gap}$ with $\Delta \approx n\Delta_{\rm
0,sh}\tilde{\Delta}$ gives the radius of the $n$'th collision,
\begin{equation}
R_{\rm col,n} \sim \fracb{\Delta}{\Delta_{\rm
0,sh}}^{2/(4-k)}R_{\rm cr,sh} \sim (n\tilde{\Delta})^{2/(4-k)}R_{\rm cr,sh}\sim
\fracb{n\tilde{\Delta}}{N}^{2/(4-k)}R_{\rm cr}\sim \fracb{n}{N}^{2/(4-k)}R_{\rm dec}\ ,
\end{equation}
keeping in mind that $R_{\rm dec}\sim\tilde{\Delta}^{2/(4-k)}R_{\rm
cr}$ in this regime.

\end{document}